\documentclass[reprint, amsmath,amssymb, aps, groupedaddress, a4paper]{revtex4-1}
\usepackage[a4paper,margin=2cm]{geometry}

\usepackage{amsmath}
\usepackage{graphicx}
\usepackage{dcolumn}
\usepackage{bm}
\usepackage[hidelinks]{hyperref}
\usepackage[mathlines]{lineno}
\usepackage{placeins}

\usepackage{xcolor}
\usepackage{siunitx}
\usepackage[version=4]{mhchem}

\newcommand{\ve}[1]{\mathbf{#1}}
\newcommand{\te}[1]{\overline{\overline{\mathbf{#1}}}}

\begin{document}

\author{Sergejs~Boroviks}
\email[]{sergejs.boroviks@epfl.ch}
\affiliation{Nanophotonics and Metrology Laboratory, Swiss Federal Institute of Technology Lausanne (EPFL), Lausanne, Switzerland}

\author{Andrei~Kiselev}
\affiliation{Nanophotonics and Metrology Laboratory, Swiss Federal Institute of Technology Lausanne (EPFL), Lausanne, Switzerland}

\author{Karim~Achouri}
\affiliation{Nanophotonics and Metrology Laboratory, Swiss Federal Institute of Technology Lausanne (EPFL), Lausanne, Switzerland}

\author{Olivier J.F.~Martin}
\email{olivier.martin@epfl.ch}
\affiliation{Nanophotonics and Metrology Laboratory, Swiss Federal Institute of Technology Lausanne (EPFL), Lausanne, Switzerland}

\date{January 30, 2023}

\title[]{Demonstration of a plasmonic nonlinear pseudo-diode}


\begin{abstract}
We demonstrate a nonlinear plasmonic metasurface that exhibits strongly asymmetric second-harmonic generation: nonlinear scattering is efficient upon excitation in one direction and it is substantially suppressed when the excitation direction is reversed, thus enabling a diode-like functionality.
A significant (approximately~\SI{10}{dB}) extinction ratio of SHG upon opposite excitations is measured experimentally and those findings are substantiated with full-wave simulations.
The combination of two commonly used metals -- aluminium and silver -- produces a material composition asymmetry that results into a bianisotropic response of the system, as confirmed by performing homogenization analysis and extracting an effective susceptibility tensor. Finally, we discuss the implications of our results from the more fundamental perspectives of reciprocity and time-reversal asymmetry.
\end{abstract}

\maketitle

High-performance nanoscale devices that allow transmission of light only in one direction -- optical isolators -- remain a long-coveted research objective for optical engineers. This problem is nontrivial due to the fundamental property of electromagnetic waves: in linear time-invariant (LTI) media and in the absence of an external time-odd bias, such as a magnetic field, they propagate reciprocally, i.e.\ the same way in the forward and backward directions. This property is linked with the time-reversal symmetry of the macroscopic Maxwell's equations and can be shown via the Lorentz reciprocity theorem, which specifically applies to LTI media~\cite{Caloz_2018, Asadchy_2020, Achouri_2021_PRB}. 
However, despite recent comprehensive publications on this topic~\cite{Jalas_2013, Sounas_2017_NatPhot,Caloz_2018, Asadchy_2020, Sigwarth_2022}, there remains a tangible confusion in the community about the difference between true nonreciprocity and deceptively similar time-reversal asymmetric response. For example, time-invariant and bias-less lossy systems may exhibit contrast upon excitation from opposite directions, but they do not qualify as optical isolators since they possess a symmetric scattering matrix and thus obey Lorentz reciprocity~\cite{Fan_2012}.
Furthermore, in the case of devices based on nonlinear effects, the distinction between true and pseudo-isolators is even more intricate. In particular, devices based on  Kerr-type nonlinearities~\cite{Cortufo_2021} are intrinsically limited by dynamic reciprocity: they can only perform as pseudo-isolators, since they do not exhibit unidirectional transmission upon simultaneous excitation from opposite directions~\cite{Shi_2015, Fernandes_2018}. One aim of this work is to explore possibilities to overcome this limitation and demonstrate how it can be turned into an advantage with an appropriate application.

In that context, photonic metasurfaces -- artificial planar materials constituted of subwavelength elements -- have been identified as a promising platform for the realization of miniature optical isolators or asymmetric devices~\cite{Shaltout_2019}.
To this end, let us highlight recent progress in the development of two classes of metasurfaces -- nonlinear and bianisotropic metasurfaces. These two classes are particularly relevant to the scope of our work, since combining their features enables realization of unconventional functionalities, such as aforementioned nonlinearly induced nonreciprocity~\cite{Menzel_2010, Mahmoud_2015, Lawrence_2018, Chen_2021, Cheng_2021}, directional harmonic generation~\cite{Yang_2017_NanoLett, Xu_2020, Nauman_2021} and nonlinear beam shaping \cite{Tymchenko_2016, BarDavid_2019}.

Nonlinear metasurfaces~\cite{Minovich_2015, Li_2017, Krasnok_2018} have the potential to replace bulky optical crystals and thus miniaturize nonlinear optical devices. 
Among other applications, plasmonic metasurfaces have proven to be interesting for second-harmonic generation (SHG)~\cite{Lee_2014, Kauranen_2012, Butet_2015}, which is a second-order nonlinear optical process in which an excitation wave with frequency $\omega$ is converted into a wave with double frequency $2\omega$~\cite{Boyd_2020}.
However, the second-order nonlinear response of plasmonic metals is weak due to  their centrosymmetric crystal structure, which is only broken at the surface, giving rise to a non-vanishing surface normal component of the second-order susceptibility tensor $\chi_{\perp \perp \perp}^{(2)}$. Yet, the overall SHG efficiency remains small due to the reduced interaction volume: essentially, the nonlinear process occurs within the few atomic layers at the metal surface, since the bulk metal is opaque for visible and infrared light and its bulk second-order response is vanishing. Nevertheless, this limitation can be partially overcome by the virtue by virtue of the field enhancement associated with surface plasmon resonances at metal surfaces.
Thus, various SHG enhancement schemes were proposed for plasmonic metasurfaces, based on multipolar resonances~\cite{Chandrasekar_2015, Kruk_2015, Smirnova_2016, Bernasconi_2016, Butet_2017, Yang_2017_ACSPhotonics, Kiselev_2019}, plasmonic lattice resonances~\cite{Gupta_2021, Abir_2022} and even light-induced centrosymmetry breaking~\cite{Li_2021}.

On the other hand, bianisotropic metasurfaces allow engineering the polarization response to realize highly efficient refraction devices through the combination of electric and magnetic effects~\cite{Asadchy_2018, Achouri_2021}. The bianisotropic response, which emerges in structures with broken spatial symmetries~\cite{Achouri_2022a}, implies that the material acquires magnetic polarization upon excitation with an electric field, and vice versa, electric polarization is produced by a magnetic field. Such a magneto-electric coupling gives rise to the spatial dispersion (i.e.\ wavevector-dependent response) that enables an excitation angle-dependent operation~\cite{Overvig_2022}. For example, in lossy systems, it may lead to asymmetric reflection and absorption, which will be discussed further in relation to our work.

\begin{figure*}[ht!]
    \centering
    \includegraphics[width=0.7\linewidth]{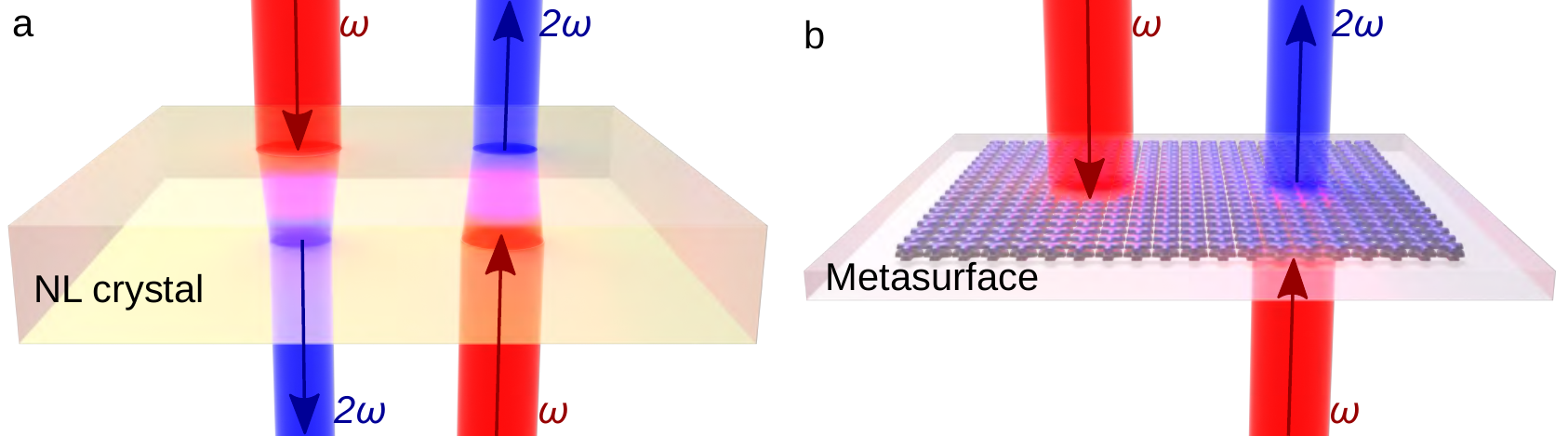}
    \caption{Comparison of conventional and asymmetric SHG: (a) symmetric SHG from a nonlinear (NL) crystal; (b) asymmetric SHG from a nonlinear bianisotropic  metasurface.
    \label{Fig1}}
\end{figure*}

In this work, we demonstrate theoretically and experimentally a plasmonic metasurface that exhibits asymmetric SHG. The operation of the device is conceptually depicted in Fig.~\ref{Fig1}: in contrast to a conventional nonlinear crystal, second-harmonic (SH) is efficiently generated only upon one excitation direction, which essentially,  enables a nonlinear optical pseudo-diode functionality (to be distinguished from optical isolators and pseudo-isolators). Such an asymmetric response imposes a structural asymmetry of the system and previously proposed theoretical designs with similar functionalities have relied on a geometric asymmetry, which might be difficult to realize experimentally~\cite{Poutrina_2016, Mobini_2021, Jin_2020, Kim_2021, Liu_2021}. Here, we take a different route and implement a structural asymmetry through the utilization of two common plasmonic meterials -- silver (\ce{Ag}) and aluminium (\ce{Al}) -- in a metasurface and show that substantial direction-dependent SHG (up to approx.\ \SI{16.9}{dB} in theory and approx.\ \SI{10}{dB} in experiment). A major advantage of this two-dimensional design is that such a material asymmetry is relatively easy to implement using standard nanofabrication techniques, e.g.\ single-exposure electron-beam lithography (EBL)~\cite{Abasahl_2021}. Furthermore, the combination of plasmonic metals is known to enhance nonlinear processes~\cite{Wang_2019, Wang_2021}.
To the best of our knowledge, this is the first experimental demonstration of a \textit{plasmonic} metasurface for asymmetric SHG, although we note that in a recent experimental demonstration Kruk et al.\ utilized a combination of dielectric nonlinear materials for third-harmonic generation~\cite{Kruk_2022}. Additionally, we perform homogenization analysis of the metasurface to extract effective susceptibilities and reveal bianiostropic property of our metasurface. Finally, we discuss the fundamental implications of our results in the context of nonreciprocity.

\begin{figure*}[htbp]
    \centering
    \includegraphics[width=\textwidth]{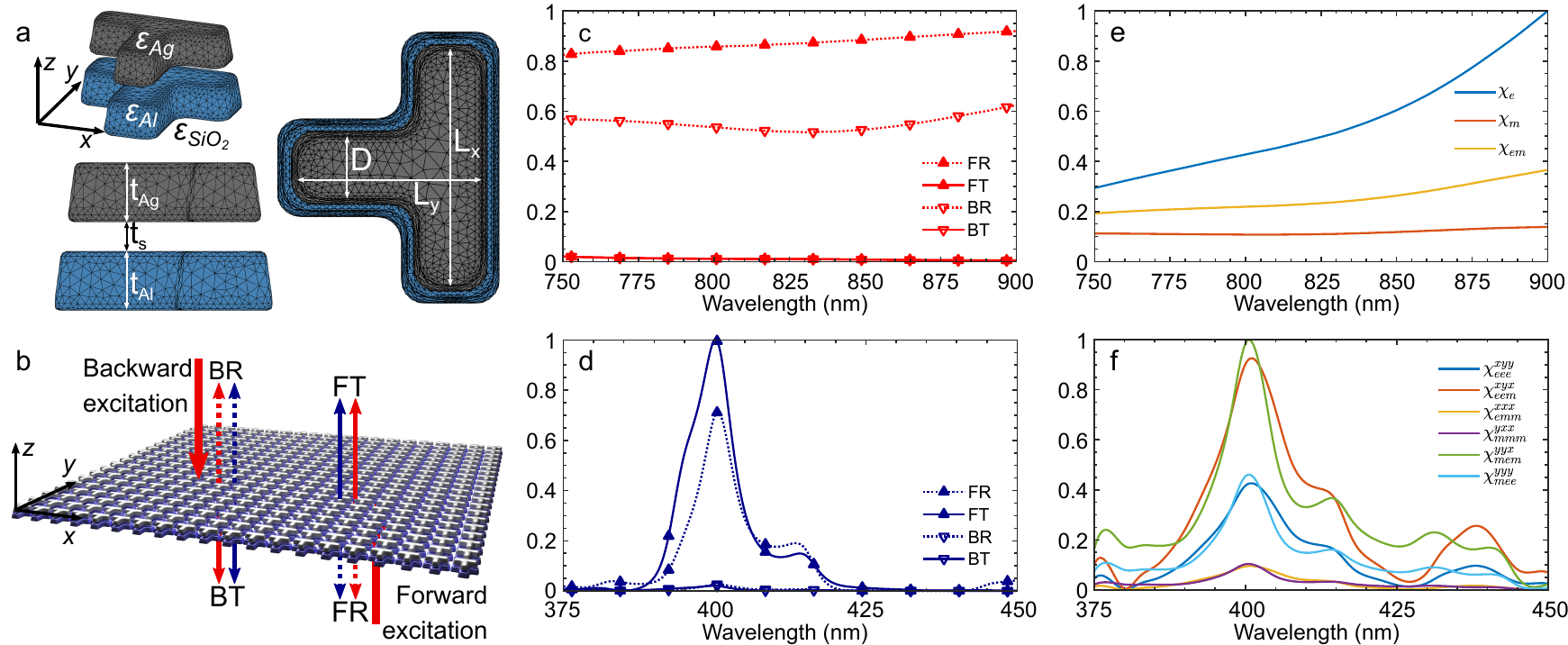}
    \caption{Design and simulated performance of the nonlinear bianisotropic metasurface. (a) Schematics of the system and the considered forward-excitation transmission (FT) and reflection (FR), as well as backward-excitation transmission (BT) and reflection (BR); thick solid red arrows indicate the excitation waves; thin solid (dashed) arrows indicate the transmitted (reflected) waves at the excitation frequency in red and at the SH frequency in blue. (b) Schematic drawing of the metasurface unit cell in isometric-, top- and side-views with indicated geometric and material parameters. Simulated metasurface reflectance and transmittance (c) in the linear regime and (d) at the SH frequency. Relevant components of the extracted (e) linear and (f) nonlinear effective susceptibility tensors.
    \label{Fig2}}
\end{figure*}

The building block of the metasurface -- the meta-atom -- is schematically depicted in Fig.~\ref{Fig2}a. 
It is comprised of two T-shaped nanostructures made of \ce{Al} and \ce{Ag} that are stacked one on top of the other and separated by a thin silicone dioxide (\ce{SiO2}) spacer. These nanostructures are embedded in \ce{SiO2} and arranged in a square lattice with the period of $\Lambda=$\SI{250}{\nano\meter}. Such a periodicity is sufficiently small to avoid diffraction in both linear and nonlinear regimes, as the metasurface is designed for the excitation with the vacuum wavelength of $\lambda_0=$\SI{800}{\nano\meter} (the effective wavelength in \ce{SiO2} is $\sim$\SI{537}{\nano\meter}) and SHG at $\lambda_\textrm{SH}=$\SI{400}{\nano\meter} ($\sim$\SI{268}{\nano\meter} in \ce{SiO2}).

As shown in Fig.~\ref{Fig2}b, we consider two different excitation conditions that are indicated with red thick arrows: forward (in the direction along the $+z$-axis) and backward (along the $-z$-axis) propagating plane waves that are $x$-polarized. In the linear regime, each of the two waves gives rise to transmitted (red solid arrows) and reflected (red dashed arrows) waves, which are labeled as forward-excited reflection (FR) and transmission (FT), or backward-excited reflection (BR) and transmission (BT).
Additionally, both excitations produce signals at the SH frequency (shown with blue arrows). For the SH signals, we use the same naming convention  as the waves produced by linear scattering, Fig.~\ref{Fig2}b. For the reflected and transmitted waves at the excitation frequency, we measure the co-polarized $x$-component of the electric field, whereas for the SHG waves, the cross-polarized $y$-component is measured, as it is found to be dominant (see Fig.~S3 in the Supporting Information).

T-shaped meta-atoms provide almost independent control of the spectral positions for the resonances both at the excitation and SH frequencies by varying the lateral dimensions $L_x$ and $L_y$~\cite{Czaplicki_2015}. As can be seen from Fig.~S1 in the Supporting Information, for a fixed wavelength, the transmission in the linear regime is tuned by varying $L_x$. In the nonlinear regime, the transmission and reflection are controlled by both $L_x$ and $L_y$. Importantly, for forward excitation, the maximum in SHG transmission coincides with the minimum in linear transmission (compare panels a and b in Fig.~S1 in the Supporting Information). The other geometric parameters $L_\textrm{s}$, $D$, $t_{\ce{Ag}}$ and $t_{\ce{Al}}$ do not have a strong influence on the resonance wavelength of the fundamental mode, however they affect the scattering cross-section of the meta-atoms via the retardation effects~\cite{Kottmann_2001}, which, in turn determines the overall transmission and SHG intensity (see Fig.~S2 in the Supporting Information). The sidewalls of the meta-atom are tilted by \SI{10}{\degree} and the edges and corners are rounded with a \SI{5}{\nano\meter} radius to mimic the experimentally fabricated structures, as discussed below.

We select $L_x=$\SI{135}{\nano\meter}, $L_y=$\SI{195}{\nano\meter}, $L_\textrm{s}=$\SI{25}{\nano\meter} and $D=t_{\ce{Ag}}=t_{\ce{Al}}=$\SI{50}{\nano\meter}, since these parameters  maximize SHG upon forward excitation at the design wavelength. Such meta-atom dimensions result in minimal transmission in the linear regime and sufficiently high extinction ratio of SHG upon forward and backward excitation (see the parametric sweeps in Fig.~S1 in the Supporting Information). Furthermore, in the $L_x$ and $L_y$ parameter space, the forward-excitation SHG peak is broad, which implies that the metasurface efficiency is weakly sensitive to  deviations from the nominal dimensions, thus easing nanofabrication tolerances.   

The simulations are performed in two steps using a custom-developed numerical electromagnetic solver based on the surface integral equation~\cite{Gallinet_2010,Butet_2013}. First, the linear fields are computed with a plane-wave excitation and periodic boundary conditions.
For the SHG simulations, the nonlinear surface polarization $P_{\perp}^{(2\omega)} = \chi^{(2)}_{\perp\perp\perp} E_\perp^{\omega} E_\perp^{\omega}$ is used as a source, where the normal components of the surface fields $E_\perp^\omega$ are obtained from the linear simulations.

The simulated reflectance and transmittance in the linear and SHG regimes are shown in Fig.~\ref{Fig2}c and d. 
In the simulations, we use interpolated values $\varepsilon_{\ce{Al}}$ and $\varepsilon_{\ce{Ag}}$ of the experimental permittivity data from McPeak et al.~\cite{McPeak_2015},  and for the permittivity of the background medium we use $\varepsilon_{\ce{SiO2}}=2.22$. Among the noble metals, \ce{Ag} is known to have the lowest losses at optical frequencies, whereas \ce{Al} has recently attracted attention as a low cost alternative plasmonic material~\cite{CastroLopez_2011,Knight_2014,Gerard_2014, Thyagarajan_2016}. Apart from its low cost, \ce{Al} is known to have the highest second-order nonlinear susceptibility among the plasmonic materials, in particular its surface normal component $\chi^{(2)}_{\perp\perp\perp}$~\cite{Krause_2004}, it also exhibits an interband transition-related absorption peak at \SI{800}{\nano\meter} (see Fig.~S4 in the Supporting Information).

As shown in Fig.~\ref{Fig2}c, in the linear regime, the transmission $T$ for both forward and backward excitations is exactly the same, as imposed by reciprocity. However, the reflection $R$ and absorption $A$, which are related to transmission as $A+R=1-T$, depend on the excitation direction, as they are not restricted by reciprocity and depend on the spatial asymmetry of the system. 
The asymmetric reflection and absorption of the system can be analyzed by considering an isolated meta-atom. As can be seen in Fig.~S5c and d in the Supporting Information, forward and backward excitations give rise to two distinct electric field distributions. In particular, the electric field concentration in the \ce{Al} part of the structure is strongly dependent on the excitation direction. Although the response is primarily dipolar for both excitations (see Fig.~S6a and b in the Supporting Information), this results in asymmetric linear scattering and absorption cross-sections, which is a characteristic of \textit{bianisotropic} systems~\cite{Cheng_2021}. In fact, it is presence of the losses that enables asymmetric scattering when the structure is illuminated from opposite directions, whereas the extinction cross-section remains exactly the same, as imposed by reciprocity~\cite{Sounas_2014}. 

In turn, the SHG response that is plotted in Fig.~\ref{Fig2}d, has an even stronger dependence on the excitation direction: both nonlinear FT and RT are more than two orders of magnitude stronger than the BT an BR at \SI{400}{\nano\meter}. A multipolar analysis of an isolated meta-atom (see Fig.~S6c and d in the Supporting Information), shows that the electric dipolar and quadruplar modes are excited more efficiently at \SI{400}{\nano\meter} upon forward excitation. This is due to the aforementioned different electric-field distributions at the surface of the T-shaped particles, that become the sources for  the SHG. 

To further elucidate the significance of bianisotropy in such an asymmetric response, we extracted the effective susceptibilities from the simulated electromagnetic fields following the previously documented procedure of metasurface homogenization analysis~\cite{Achouri_2017, Achouri_2018, Achouri_2022}.
Briefly, the expressions for nonlinear susceptibilities are derived from the generalized sheet transition conditions and are calculated using the simulated reflected and transmitted fields upon different excitation conditions at $\omega$ and $2\omega$ frequencies.

In Fig.~\ref{Fig2}e and f we plot the extracted effective susceptibility tensor elements that are relevant to the considered excitation conditions. For both linear and nonlinear susceptibilities, the magneto-electric coupling (corresponding to the terms with mixed ``$\textrm{e}$" and ``$\textrm{m}$" subscripts in Fig.~\ref{Fig2}e and f) is non-negligible. 
The asymmetric response becomes apparent by noting that the induced linear and nonlinear polarizations are given by
\begin{widetext}
    \begin{subequations}
        \begin{gather}
            \ve{P}^{\omega} = \te{\chi}_\textrm{ee}^{\:\omega} \cdot \ve{E}^{\omega}  + \te{\chi}_\textrm{em}^{\:\omega} \cdot \ve{H}^{\omega},\\
            \ve{P}^{2\omega} = \te{\chi}_\textrm{ee}^{\:2\omega} \cdot \ve{E}^{2\omega}  + \te{\chi}_\textrm{em}^{\:2\omega} \cdot \ve{H}^{2\omega} + \te{\chi}_\textrm{eee}^{\:\omega}:\ve{E}^{\omega}\ve{E}^{\omega} + \te{\chi}_\textrm{eem}^{\:\omega}:\ve{E}^{\omega}\ve{H}^{\omega} + \te{\chi}_\textrm{emm}^{\:\omega}:\ve{H}^{\omega}\ve{H}^{\omega}.
        \end{gather}    
    \end{subequations}
\end{widetext}

In the linear regime, the non-negligible magneto-electric coupling term $\chi_\textrm{me}$ results in an asymmetric absorption and reflection. 
As for the nonlinear effective susceptibility tensors, the dominant components are $\chi_\textrm{mem}^{yyx}$ and $\chi_\text{eem}^{xyx}$, which relate magnetic/electric excitations with electric/magnetic responses along orthogonal directions and result in strongly asymmetric SHG.

\begin{figure*}[ht!]
    \centering
    \includegraphics[width=\textwidth]{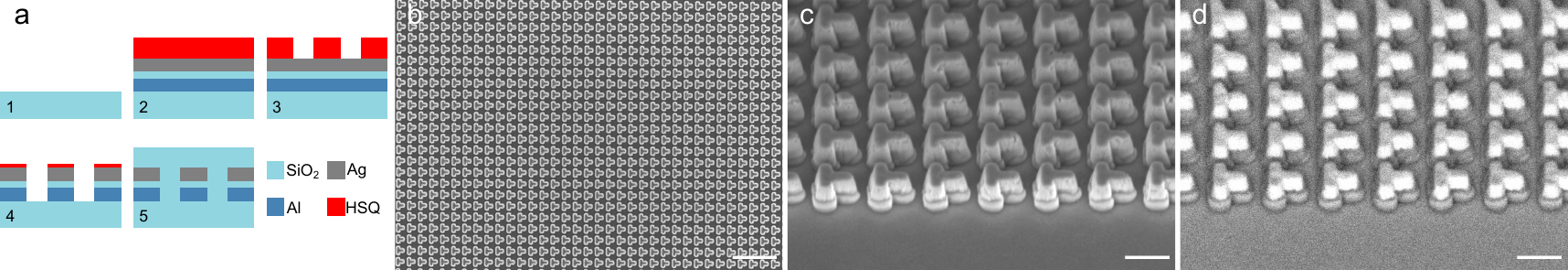}
    \caption{Fabrication of the bimetallic metasurface. (a) Flowchart of the fabrication: 1. Initial substrate; 2. \ce{Al}, \ce{SiO2}, \ce{Ag} and HSQ thin films deposition; 3. E-beam exposure; 4. IBE; 5. Covering with a thick \ce{SiO2} film. SEM images of the fabricated structure acquired using different detectors and tilt angles: (b) top view SE (scale bar: \SI{1}{\micro\meter}. (c) \SI{45}{\degree}-tilted view SE (scale bar: \SI{200}{\nano\meter}. (d) \SI{45}{\degree}-tilted view BSE (scale bar: \SI{200}{\nano\meter}. 
    \label{Fig3}}
\end{figure*}

To verify  experimentally this asymmetric nonlinear response, we fabricated and characterized a metasurface device.
Instead of the widespread lift-off process, we employ the ion beam etching (IBE) technique which enables the fabrication of stratified nanostructures, in particular metal-dielectric composites, with sharper features~\cite{Ray_2020,Abasahl_2021}. The schematic flowchart of the fabrication process is shown in Fig.~\ref{Fig3}a. We use a \SI{150}{\micro\meter}-thick D 263 glass wafer (Schott) which is coated with \SI{50}{\nano\meter}-thick \ce{Al} and \SI{25}{\nano\meter} \ce{SiO2} films using RF sputtering (Pfeiffer SPIDER 600). Next, we deposit a \SI{50}{\nano\meter} thick \ce{Ag} layer using an e-beam assisted evaporator (Alliance-Concept EVA 760). The T-shaped pattern arrays are exposed in the hydrogen silsesquioxane (HSQ, XR-1541-006 from DuPont), which is a negative tone e-beam resist, using electron beam lithography (Raith EBPG5000+). The formation of the exposed patterns in the thin films is performed using a low-power argon IBE (Veeco Nexus IBE350, operated at a \SI{300}{\volt} acceleration voltage). An important point for this last step is the pulsed IBE operation: 10 s of etching followed by 30 s of cooling to avoid damaging the sample by substrate overheating. The typical overall IBE process time is \SI{160}{s}, and the etching depth is controlled in-situ using a mass-spectrometer, which allows real-time monitoring of the etched material composition: the etching process is stopped as soon as the \ce{Al} flux drops to a minimum.
The fabrication results are shown in the scanning electron microscope (SEM) images in Fig.~\ref{Fig3}b-d. The morphology of the fabricated structure can be inspected in Fig.~\ref{Fig3}c: intrinsically, the IBE process results in tilted sidewalls (approx. \SI{10}{\degree}) and rounded corners and edges. Although such features are typically undesired, they are not expected to degrade the performance of the metasurface, as these were taken into account in the simulations.  In turn, the layered material composition can be well identified in the image acquired with the back-scattered electron (BSE) detector in Fig.~\ref{Fig3}d. 
In the last fabrication step, we cover the metallic nanostructures with a thick \ce{SiO2} layer (approx. \SI{300}{\nano\meter}) which serves two purposes: it acts as a protective layer preventing degradation of the \ce{Al} and \ce{Ag} nanostructures, and simplifies the physical conditions by having identical permittivities above and below the metasurface. 

\begin{figure}[hbp!]
    \centering
    \includegraphics[width=0.9\linewidth]{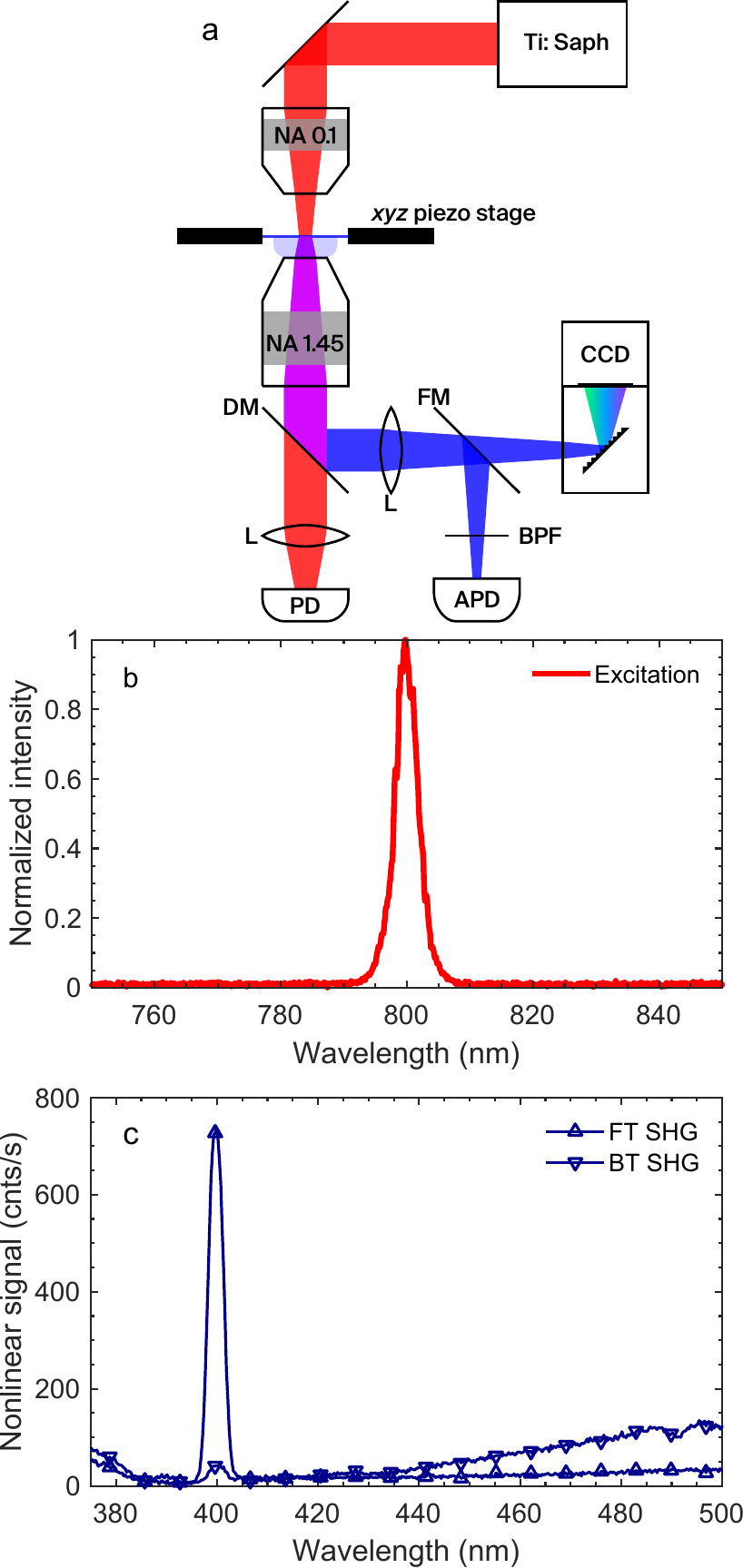}
    \caption{Optical characterization of the metasurface: (a) measurement setup. (b) Excitation spectrum. (c) Nonlinear spectra.
    \label{Fig4}}
\end{figure}

The experimental setup and the results for the optical characterization of the fabricated sample are shown in Fig.~\ref{Fig4}. As an excitation light source, we use a mode-locked Ti:Saph laser that outputs approx. \SI{120}{\femto\second} pulses with a central wavelength of \SI{800}{\nano\meter}. The excitation light is weakly focused onto the metasurface with a low magnification objective ($\text{NA}=0.1$), which results in a focal spot with a \SI{10}{\micro\meter} FWHM mimicking the plane wave excitation used in the simulations. The spectrum of the nonlinearly generated light is shown in Fig.~\ref{Fig3}a. Apart from the characteristic SHG peak at \SI{400}{nm}, it has a tail at longer wavelengths, which is attributed to nonlinear photo-luminescence (NPL).

As an interesting side-effect, we note that the NPL signal is substantially larger for BT than for FT. This fact can be explained by the peculiarity of the two-photon absorption mechanism in metals that induces the NPL. 
As opposed to the coherent nature of two-photon absorption in molecules or dielectrics, in metals it can be regarded as a cascaded process. Specifically, two photons are absorbed sequentially rather than simultaneously~\cite{Beversluis_2003, Muhlschlegel_2005, Biagioni_2009}. Absorption of the first photon gives rise to an intraband transition in the conduction band and creates a vacancy below the Fermi level. Thus, the second photon results in an interband transition that fills the vacancy in the conduction band and creates one in the valence band. Both of these photon absorption steps are linear, but result in an effective nonlinearity. Thus, higher linear absorption upon backward excitation (see Fig.~S5 and discussion above), results in a higher probability of two-photon absorption and subsequent NPL, which is consistent with our observations.

Such asymmetric behaviour is sometimes referred to as \textit{"nonreciprocal SHG"}, both in the metasurfaces~\cite{Poutrina_2016} and solid state physics~\cite{Toyoda_2021, Mund_2021} communities. We share the view that such a nomenclature is improper in the case of SHG, since the concept of  nonrecipocity is not well-defined for nonlinear optics~\cite{Trzeciecki_2000, Sounas_2017_PRL, Achouri_2018}. For any $N$-port system, the Lorentz reciprocity implies the symmetry of the scattering matrix $\te{S}^T=\te{S}$, where $^T$ denotes the transpose operator. 
In the case of a two-port system like that considered in this work in the linear regime, the scattering matrix is given by
\begin{equation}
    \te{S}=\begin{bmatrix}
        S_{11} & S_{12}\\ S_{21} & S_{22}
    \end{bmatrix},
\end{equation}
and  reciprocity requires that the transmission coefficients $S_{12}$ and $S_{21}$ are equal. 
However, it does not impose any limitations on the reflection coefficients $S_{11}$ and $S_{22}$.
This is true for our system in the linear regime, since the transmissions for forward and backward excitations are equal, while the reflections are asymmetric. 

However, in the nonlinear regime, our metasurface cannot be regarded as a two-port system anymore, since the SH emission represents a distinct electromagnetic mode. Therefore, this system must be at least considered as a 4-port system (assuming that higher-order harmonic generation is negligible), represented with the following scattering matrix:
\begin{multline}
         \te{S}=\begin{bmatrix}
       \te{S}^{\omega{\scriptscriptstyle\rightarrow}\omega} & \te{S}^{2\omega{\scriptscriptstyle\rightarrow}\omega} \\
       \te{S}^{\omega{\scriptscriptstyle\rightarrow}2\omega} & \te{S}^{2\omega{\scriptscriptstyle\rightarrow}2\omega}
  \end{bmatrix}\\=
  \begin{bmatrix}
        S_{11}^{\omega{\scriptscriptstyle\rightarrow}\omega} & S_{12}^{\omega{\scriptscriptstyle\rightarrow}\omega} & S_{11}^{2\omega{\scriptscriptstyle\rightarrow}\omega}  & S_{12}^{2\omega{\scriptscriptstyle\rightarrow}\omega} \\
        S_{21}^{\omega{\scriptscriptstyle\rightarrow}\omega} & S_{22}^{\omega{\scriptscriptstyle\rightarrow}\omega} & S_{21}^{2\omega{\scriptscriptstyle\rightarrow}\omega}  & S_{22}^{2\omega{\scriptscriptstyle\rightarrow}\omega} \\
        S_{11}^{\omega{\scriptscriptstyle\rightarrow}2\omega} & S_{12}^{\omega{\scriptscriptstyle\rightarrow}2\omega} & S_{11}^{2\omega{\scriptscriptstyle\rightarrow}2\omega}  & S_{12}^{2\omega{\scriptscriptstyle\rightarrow}2\omega} \\
        S_{21}^{\omega{\scriptscriptstyle\rightarrow}2\omega} & S_{22}^{\omega{\scriptscriptstyle\rightarrow}2\omega} & S_{21}^{2\omega{\scriptscriptstyle\rightarrow}2\omega}  & S_{22}^{2\omega{\scriptscriptstyle\rightarrow}2\omega} 
    \end{bmatrix},  
\end{multline}

which describes both linear transmission/reflection at frequencies $\omega$ and $2\omega$, as well as nonlinear processes $\omega{\scriptscriptstyle\rightarrow}2\omega$ and $2\omega{\scriptscriptstyle\rightarrow}\omega$.

In our experiment, we do not directly probe $S_{21}^{\omega{\scriptscriptstyle\rightarrow}2\omega}\stackrel{?}{=}S_{12}^{2\omega{\scriptscriptstyle\rightarrow}\omega}$, where $S_{12}^{2\omega{\scriptscriptstyle\rightarrow}\omega}$ parameters corresponds to the excitation at SH frequency and generation of a wave at frequency $\omega$. In fact, this process is known as known as parametric down conversion and it has an extremely low efficiency in comparison with SHG~\cite{Boyd_2020}. Probing this equality, as well as equality of 8 other parameters that are flipped by the $\te{S}^T$ operation, namely $S_{21}^{\omega{\scriptscriptstyle\rightarrow}\omega}\stackrel{?}{=}S_{12}^{\omega{\scriptscriptstyle\rightarrow}\omega}$, $S_{11}^{\omega{\scriptscriptstyle\rightarrow}2\omega}\stackrel{?}{=}S_{11}^{2\omega{\scriptscriptstyle\rightarrow}\omega}$,
$S_{12}^{\omega{\scriptscriptstyle\rightarrow}2\omega}\stackrel{?}{=}S_{21}^{2\omega{\scriptscriptstyle\rightarrow}\omega}$,
$S_{22}^{\omega{\scriptscriptstyle\rightarrow}2\omega}\stackrel{?}{=}S_{22}^{2\omega{\scriptscriptstyle\rightarrow}\omega}$ and
$S_{21}^{2\omega{\scriptscriptstyle\rightarrow}2\omega}\stackrel{?}{=}S_{12}^{2\omega{\scriptscriptstyle\rightarrow}2\omega}$ stand for a true reciprocity test in a four-port system.
Instead, within our experiment we show that $S_{21}^{\omega{\scriptscriptstyle\rightarrow}2\omega}\neq S_{12}^{\omega{\scriptscriptstyle\rightarrow}2\omega}$, which corresponds to an asymmetric nonlinear scattering process that is reciprocal. 
Yet, a rigorous probing of reciprocity in a nonlinear system would require sophisticated experiments that involve simultaneous excitation with the two waves at frequencies $\omega$ and $2\omega$ and precise control over their amplitude and phase~\cite{Trzeciecki_2000}. 
Nevertheless, we assert that our device essentially functions as a nonlinear optical pseudo-diode, allowing the transmission of SH signal only in one direction, which is a desired functionality for various signal processing applications~\cite{Willner_2014}.

In summary, we have demonstrated that strongly asymmetric SHG can be achieved in a plasmonic metasurface that is comprised of two common plasmonic metals -- aluminium and silver. The structural asymmetry created by the material contrast results in a strong dependence on the excitation direction, with an extinction ratio of approx.\ \SI{16.9}{dB} in theory and approx.~\SI{10}{dB} in the experiment. 
We anticipate that our findings can pave the way for further developments in the field of nonlinear bianisotropic and nonreciprocal devices, as well as inspire novel plasmonic devices with unrivaled functionalities.

\section*{acknowledgement}
The authors thank Christian Santschi and Zdenek Benes for their valuable advises on nanofabrication.\\
Funding from the Swiss National Science Foundation (grant PZ00P2\_193221) is gratefully acknowledged.

\bibliography{rfrncs}

\appendix
\onecolumngrid
\pagebreak
\section*{Supporting Information}
\renewcommand\thefigure{S\arabic{figure}}    
\setcounter{figure}{0}

\begin{figure}[!htpb]
    \centering
    \includegraphics[width=0.87\linewidth]{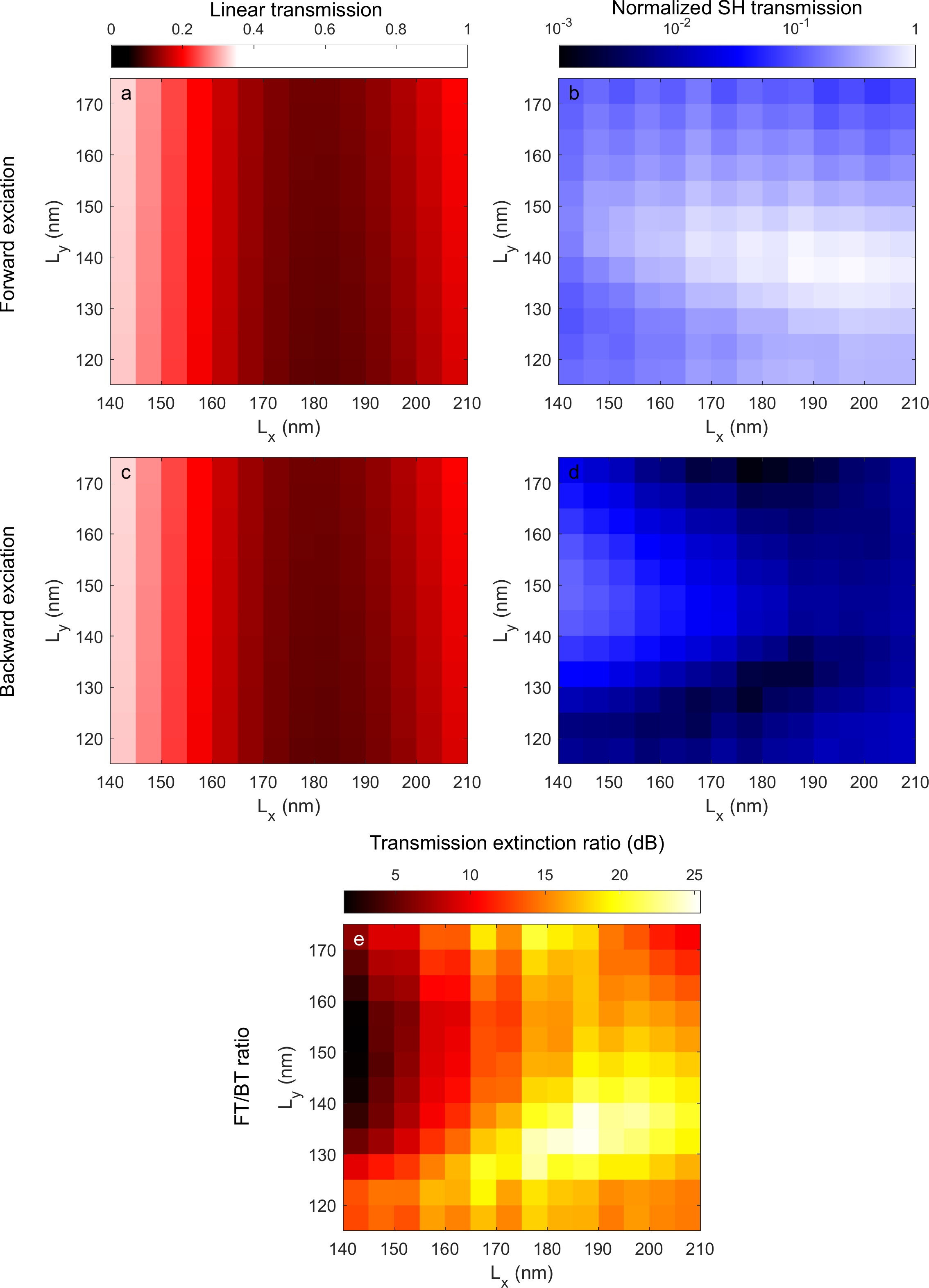}
    \caption{Control over linear and SH transmission via $L_x$ and $L_y$ geometrical parameters at excitation wavelength $\lambda_0=\SI{800}{\nano\meter}$. Other parameters are fixed: $D=t_{\ce{Ag}}=t_{\ce{Al}}=\SI{50}{\nano\meter}$ and $L_{\textrm{s}}=\SI{25}{nm}$. (a) Linear and (b) SH transmission upon forward excitation; (c) linear and (d) SH transmission upon backward excitation. (e) Forward/backward-excitation SH transmission extinction ratio. 
    \label{FigS1}}
\end{figure}

\begin{figure}[!htpb]
    \centering
    \includegraphics[width=\linewidth]{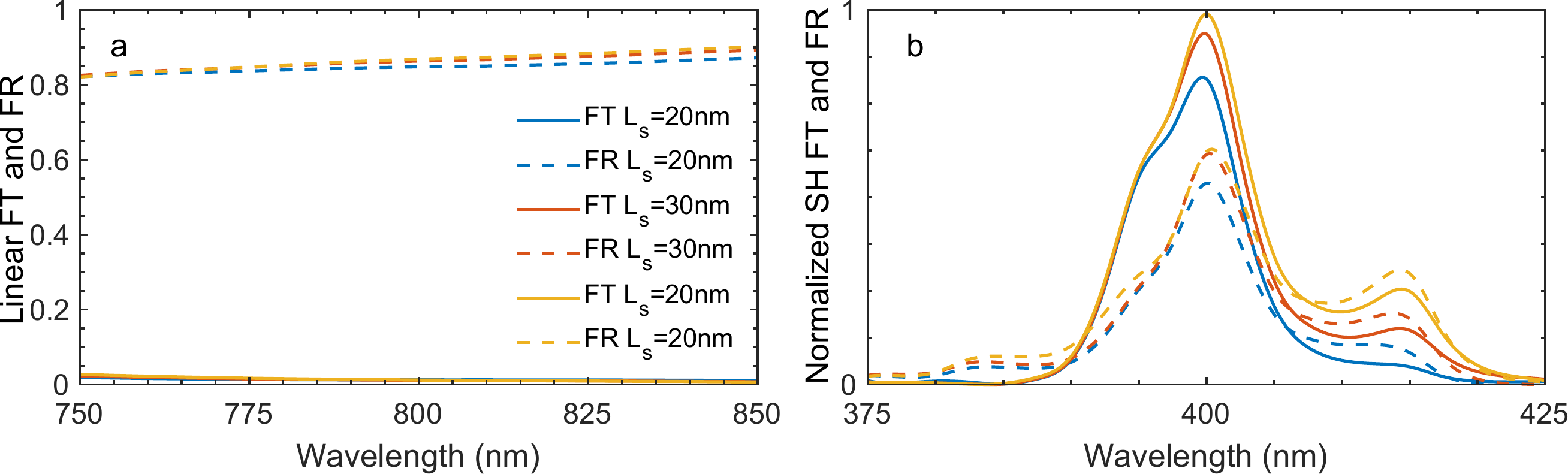}
    \caption{Dependence of the linear and SH transmission and reflection upon forward excitaiton (FT and FR) on the geometrical parameters (a) Linear  and (b) SH. Other geometrical parameters are fixed: $L_x=$\SI{135}{\nano\meter}, $L_y=$\SI{195}{\nano\meter} and $D=t_{\ce{Ag}}=t_{\ce{Al}}=\SI{50}{\nano\meter}$.
    \label{FigS2}}
\end{figure}

\begin{figure}[!htpb]
    \centering
    \includegraphics[width=\linewidth]{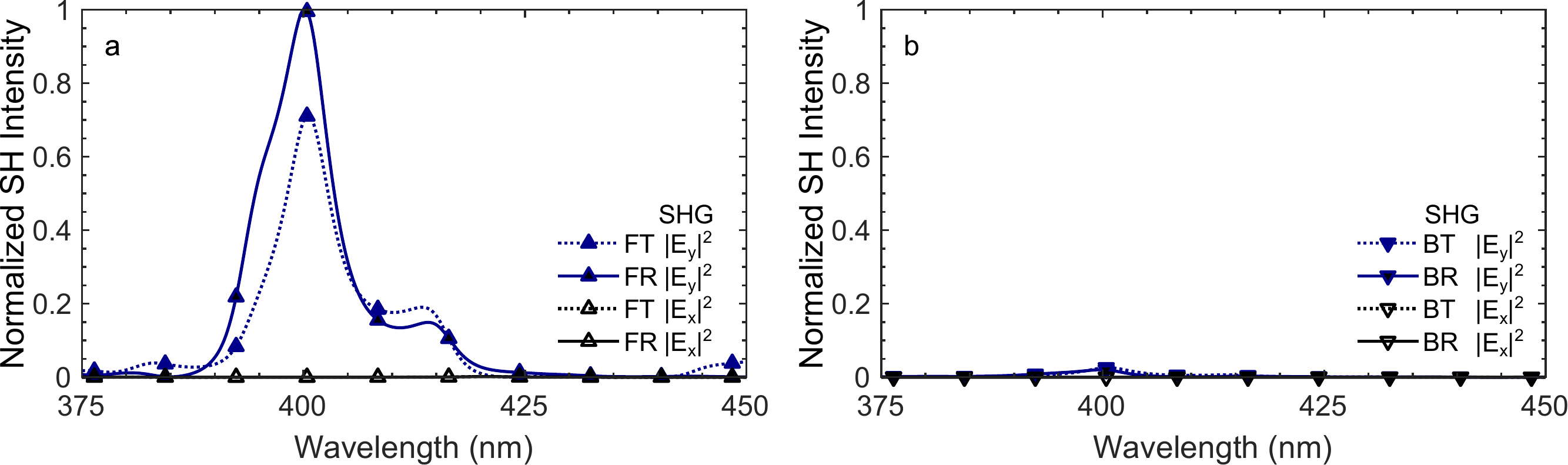}
    \caption{Electric field  components calculated for  the reflected and transmitted SH waves. In both cases, (a) forward excitation (FE) and (b) backward excitation (BE) the $E_y^{2\omega}$ component (that is orthogonal to the excitation  field $E_x^{\omega}$) is dominant.
    \label{FigS3}}
\end{figure}

\begin{figure}[!htpb]
    \centering
    \includegraphics[width=0.5\linewidth]{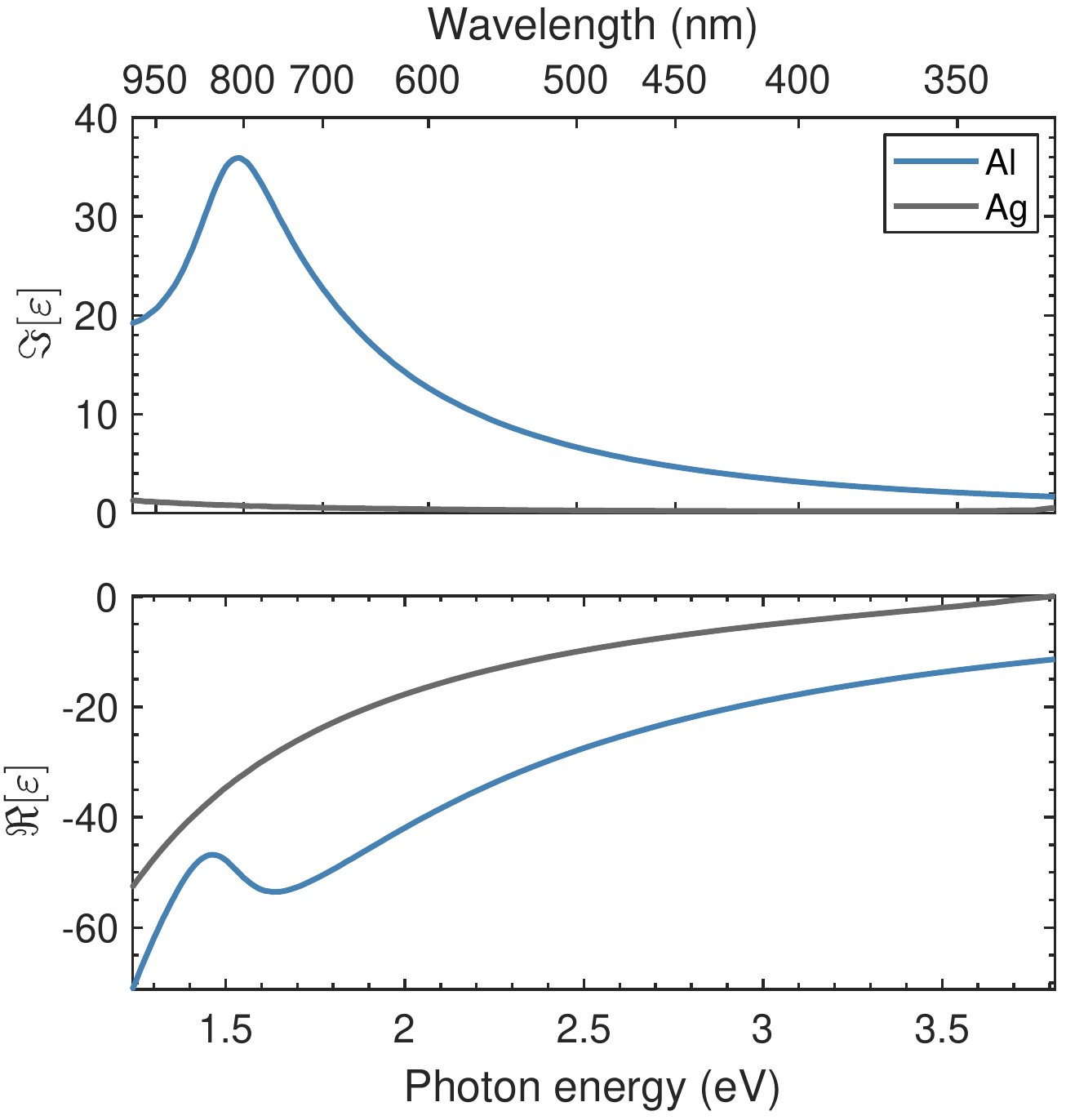}
    \caption{Dielectric permittivity of \ce{Al} (blue lines) and \ce{Ag} (gray lines) used in the simulations: real (bottom panel) and imaginary (top panel) parts of the interpolated experimental data from ref. \cite{McPeak_2015}.
    \label{FigS4}}
\end{figure}

\begin{figure}[!htpb]
    \centering
    \includegraphics[width=0.5\linewidth]{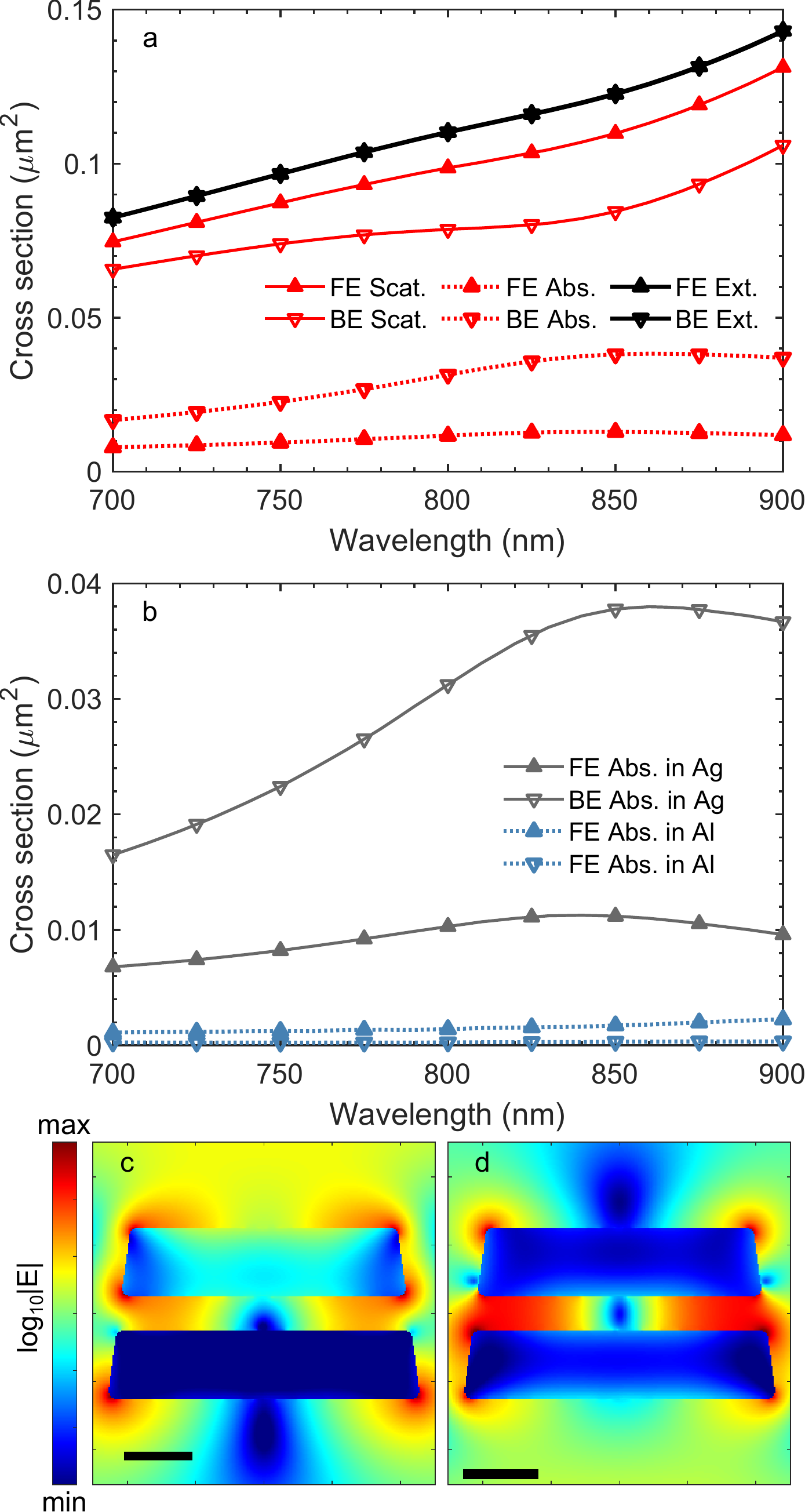}
    \caption{Isolated meta-atom linear scattering and absorption analysis. (a) Scattering (red solid lines), absorption (red dotted lines) and extinction (black-solid lines) cross-sections upon forward excitation (upward triangles) and backward excitation (downward triangles); (b) absoption in \ce{Ag} (gray solid lines) and \ce{Al} (blue dotted lines) domains upon forward excitation (upward triangles) and backward excitation (downward triangles). Pseud-color images of electric field distribution upon forward (c) and backward (d) excitations. Normalized magnitude of the electric field on the logarithmic scale is plotted.
    \label{FigS5}}
\end{figure}

\begin{figure}[!htpb]
    \centering
    \includegraphics[width=\linewidth]{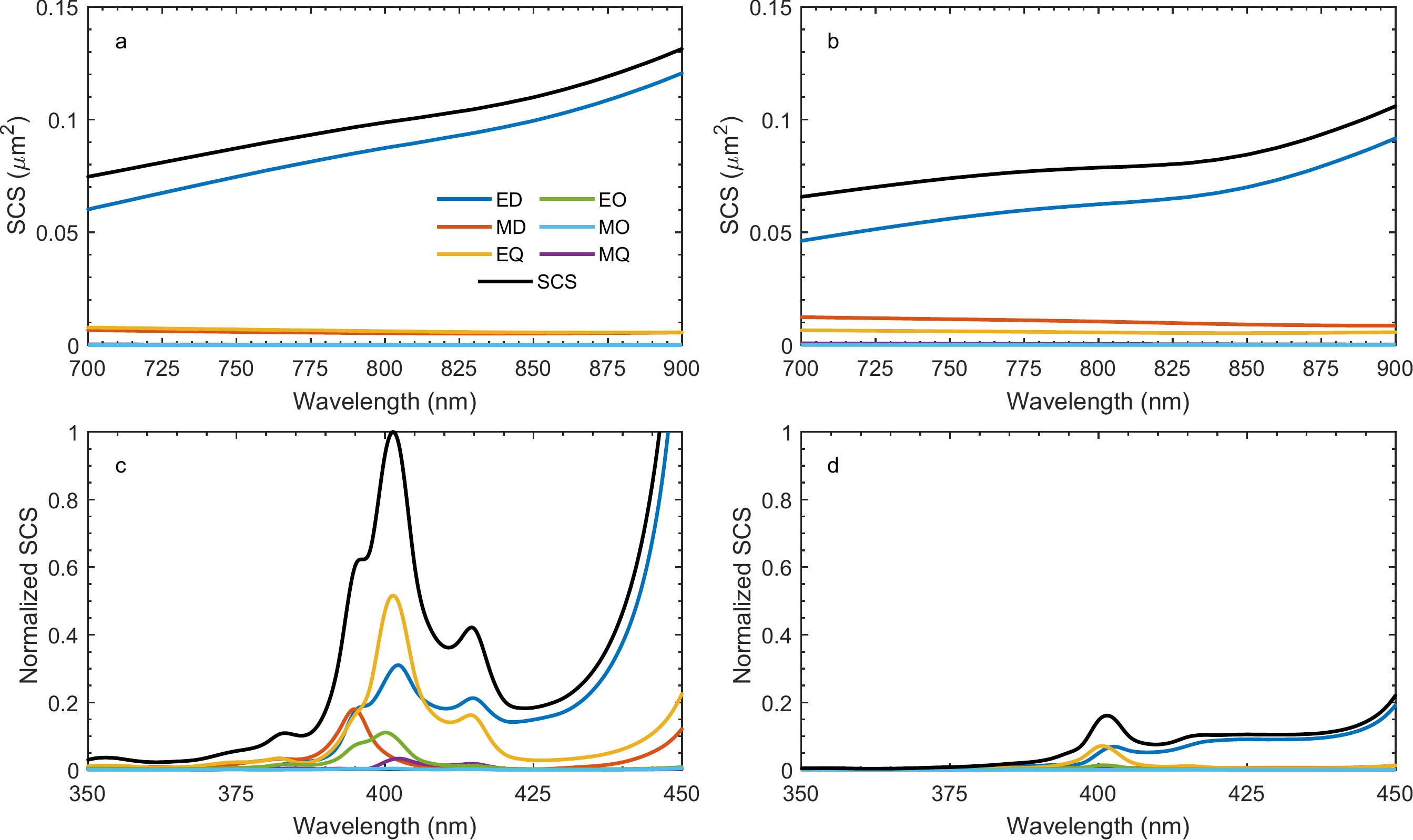}
    \caption{Isolated meta-atom multipole analysis. Vector spherical harmonic decomposition of (a) linear scattering upon forward excitation; (b) linear scattering upon backward excitation; (c) SHG scattering upon forward excitation; (d) SHG scattering upon backward excitation.
    \label{FigS6}}
\end{figure}

\end{document}